%% file: EXO-16-023_temp.tex
\pdfoutput=1

\documentclass[11pt,twoside,a4paper,cmspaper,final,collab]{cms-tdr}
\def\svnVersion{419817P}\def\cmsCernNoTag{CERN-EP-2017-025}\def\cmsCernDate{\today}\def\cmsMessage{Published in the Journal of High Energy Physics as \href{http://dx.doi.org/10.1007/JHEP07(2017)121}{\doi{10.1007/JHEP07(2017)121.}}}

\begin{document}\cmsNoteHeader{EXO-16-023}

\hyphenation{had-ron-i-za-tion}
\hyphenation{cal-or-i-me-ter}
\hyphenation{de-vices}

\RCS$Revision: 419796 $
\RCS$HeadURL: svn+ssh://svn.cern.ch/reps/tdr2/papers/EXO-16-023/trunk/EXO-16-023.tex $
\RCS$Id: EXO-16-023.tex 419796 2017-08-05 10:45:37Z abdollah $
\newlength\cmsFigWidth
\setlength\cmsFigWidth{0.48\textwidth}

\newcommand{\etau}{\ensuremath{\Pe\tauh}\xspace}
\newcommand{\mutau}{\ensuremath{\Pgm\tauh}\xspace}
\newcommand{\RZ}{\ensuremath{\PZ_{\mathrm{R}}}\xspace}
\newcommand{\ST}{\ensuremath{S_{\mathrm{T}}}\xspace}
\newcommand{\Nt}{\ensuremath{\cmsSymbolFace{N}_{\tau}}\xspace}
\newcommand{\Ne}{\ensuremath{\cmsSymbolFace{N}_{\Pe}}\xspace}
\newcommand{\Nm}{\ensuremath{\cmsSymbolFace{N}_{\PGm}}\xspace}
\newcommand{\NR}{\ensuremath{\cmsSymbolFace{N}_\mathrm{R}}\xspace}
\providecommand{\PWRpr}{\ensuremath{\cmsSymbolFace{W}_\mathrm{R}^*}\xspace}

\cmsNoteHeader{EXO-16-023}
\title{Search for third-generation scalar leptoquarks and heavy right-handed neutrinos in final states with two tau leptons and two jets in proton-proton collisions at $\sqrt{s}=13\TeV$}

\date{\today}

\abstract{
A search is performed for third-generation scalar leptoquarks and heavy right-handed neutrinos
in events containing one electron or muon, one hadronically decaying $\tau$ lepton,
and at least two jets, using a
$\sqrt{s}=13\TeV$ pp collision data sample corresponding to an integrated luminosity of
12.9\fbinv collected with the CMS detector
at the LHC in 2016. The number of observed events is found to be in
agreement with the standard model prediction. A limit is set at 95\% confidence level
on the product of the leptoquark pair production cross section and $\beta^2$, where
$\beta$ is the branching fraction of leptoquark decay to a $\tau$ lepton and
a bottom quark. Assuming $\beta=1$, third-generation leptoquarks with masses
below 850\GeV are excluded at 95\% confidence level.
An additional search based on the same event topology involves heavy right-handed neutrinos, ${\NR}$, and right-handed W bosons, ${\PWR}$, arising in a left-right symmetric extension of the standard model. In this search, ${\PWR}$~bosons are assumed to decay to a tau lepton and ${\NR}$ followed by the decay of the ${\NR}$ to a tau lepton and an off-shell ${\PWR}$ boson.
Assuming the mass of the right-handed neutrino to be half of the mass of the right-handed W boson,
$\PWR$ boson masses below 2.9\TeV are excluded at 95\% confidence level. These results improve on the limits from previous searches for third-generation leptoquarks and heavy right-handed neutrinos
with $\tau$ leptons in the final state.
}

\hypersetup{%
pdfauthor={CMS Collaboration},%
pdftitle={Search for the third-generation scalar leptoquarks and heavy right-handed neutrinos in  final states with two tau  leptons and two jets in proton-proton collisions at sqrt(s) = 13 TeV},%
pdfsubject={CMS},%
pdfkeywords={CMS, physics, leptoquarks}}

\maketitle

\section{Introduction}

A number of extensions of the standard model (SM) have been proposed that predict an enhanced production rate for events containing pairs of quarks and pairs of third-generation leptons.
One such theoretical proposal involves the existence of particles called leptoquarks (LQs), which carry color charge, fractional electric charge, and both lepton and baryon quantum numbers. The LQs arise in many models, including grand unified theories~\cite{GUT}, compositeness models~\cite{COMP1,COMP2}, and
superstring theories~\cite{SUPERSTR}.
If LQs exist, they will decay into a lepton and a quark.
At the CERN LHC, LQ pairs are predominantly produced via gluon-gluon fusion and quark-antiquark annihilation. Based on the latest experimental constraints reviewed in~\cite{Dorsner:2016wpm}, we assume that contribution of t-channel production of LQ pairs involving Yukawa coupling of a LQ, a lepton, and a quark, is small and neglected in this analysis and the main free parameter is the mass of LQ.
However, the branching fraction for the decay of a LQ into a quark and a charged lepton,
$\beta$, depends on the details of the model under consideration.
In this analysis we focus on the decay of a pair of third-generation LQs resulting in two $\tau$ leptons and two jets originating from b quark fragmentation.

A similar final state is expected in theories that postulate that the masses of the familiar left-handed neutrinos arise not from the Higgs field, but from a mechanism that involves the existence of right-handed neutrinos. One of the appealing features of left-right (L-R) symmetric extensions~\cite{Mohapatra:1974gc} of the SM is that these models predict the existence of new heavy charged (\PWR) and neutral (\RZ) gauge bosons that could be produced at LHC energies.
Heavy neutrinos
($\Ne, \Nm, \Nt$) naturally arise as the right-handed (RH)
partners of the SM neutrinos in these L-R extensions through the
see-saw mechanism~\cite{Keung:1983uu}.

In this paper, we search for these two processes by selecting final states containing two $\tau$ leptons and two jets originating from the hadronization of quarks. A search for pair production of third-generation scalar LQs is pursued by looking for events containing two $\tau$ leptons and two b quarks.
We also search for the production of a $\PWR$ boson from quark-antiquark annihilation. A heavy right-handed neutrino is produced from the decay of the $\PWR$ boson following the decay chain $\PWR\to\tau+\Nt$, where $\Nt\to\tau+\PWRpr\to\tau+\qqbar$.
In both searches, we focus on signatures with one of the $\tau$ leptons decaying into an electron or a muon, referred to as a leptonic decay $\tau_{\ell}$ in the following, and the other $\tau$ lepton decaying hadronically, denoted by \tauh.

Previous searches for third-generation LQs have been carried out
at pp, $\Pp\PAp$, $\Pep\Pem$, and ep
colliders and the most recent results are given in~\cite{ATLAS:2013oea,Khachatryan:2014ura} and references therein.
The most stringent lower limit on the mass of scalar third-generation LQs  to date, based on the final state with two $\tau$ leptons and two b jets and assuming $\beta$ = 1, is 740\GeV at 95\% confidence level (CL), from the CMS experiment~\cite{CMS-PAPERS-EXO-12-002,CMS-PAS-EXO-16-016}.
Previous searches for heavy neutrinos have been performed at LEP~\cite{Abreu:1996pa,Adriani:1992pq}, excluding heavy neutrino masses below approximately 100\GeV.
Further searches at LHC have been performed in the dielectron and dimuon channels and have excluded $\PWR$ bosons with mass up to 3\TeV using data taken at 7\TeV~\cite{ATLAS:2012ak} and at 8\TeV~\cite{Khachatryan:2014dka}.
 Using 2.1\fbinv of data of 13\TeV pp collisions collected in 2015, the CMS experiment searched for heavy neutrinos and right-handed charged bosons using events in which both $\tau$ leptons decay hadronically. That analysis excluded $\PWR$ bosons with masses below 2.35\,(1.63)\TeV at 95\% CL, assuming the \Nt mass is 0.8 (0.2) times the mass of $\PWR$ boson~\cite{CMS-PAS-EXO-16-016}. In the present search, we use a $\sqrt{\mathrm{s}}=13$\TeV pp collision data sample corresponding to an integrated luminosity of 12.9\fbinv collected with the CMS detector in 2016.

\section{The CMS detector and Monte Carlo event samples}

The central feature of the CMS apparatus is a superconducting
solenoid of 6\unit{m} internal diameter, providing a
magnetic field of 3.8\unit{T}. Within the solenoid volume
are a silicon pixel and strip tracker,
a lead tungstate crystal electromagnetic calorimeter (ECAL),
and a brass and scintillator hadron calorimeter,
each composed of a barrel and two endcap sections. Extensive forward
calorimetry complements the coverage provided by the barrel
and endcap detectors. Muons are
detected in gas-ionisation detectors embedded in the steel
flux-return yoke outside the solenoid.
A detailed description of the CMS detector, together with a
definition of the coordinate system used and the relevant
kinematic variables, can be found in Ref.~\cite{Chatrchyan:2008zzk}.

The first level of the CMS triggering system,
composed of custom hardware processors, uses
information from the calorimeters and the muon detectors to
select the most interesting events in a fixed time interval
of less than 4\mus. The high-level trigger processor
farm further decreases the event rate from around 100\unit{kHz}
to less than 1\unit{kHz}.

Background and signal processes are modeled using the
following simulated samples.
The \PYTHIA v8.205 generator~\cite{Sjostrand:8.2}
is used to model the signal and diboson (\PW{}\PW, \PW{}\PZ, and ZZ)  processes.
The LQ signal samples are generated with LQ masses ranging from
250 to 1500\GeV in steps of 50\GeV.
The branching fraction of the LQ to a $\tau$ lepton and
a bottom quark is assumed to be 100\%.
The signal samples are normalized to the next-to-next-to-leading order~\cite{Kramer:2004df,Kramer:1997hh}.
The $\PWR$ signal samples are generated with $\PWR$ boson masses
ranging from 1000 to 4000\GeV in steps of 500\GeV and the cross sections are computed in Ref.~\cite{delAguila:2007qnc}.
The \MADGRAPH v5.1.5 generator~\cite{MadGraph} is used to model
\PW+jets and \cPZ+jets processes. Single top production and \ttbar process are modelled with the \POWHEG2.0~\cite{Nason:2004rx,Frixione:2007vw,Alioli:2010xd}
generator.
The NNPDF 3.0~\cite{Ball:2014uwa} Parton Distribution Functions (PDF) are used, and all simulated samples are interfaced with \PYTHIA with the CUETP8M1 tune~\cite{Khachatryan:2015pea} to describe parton showering and hadronization.
Additional inelastic pp interactions (pileup) generated by {\PYTHIA} are overlaid on all simulated events, according to the luminosity profile of the analyzed data.
All the generated signal and background samples are processed with the simulation of the CMS detector based on  \GEANTfour\cite{Agostinelli:2002hh}.
Small differences between data and simulation in trigger,
in particle identification and isolation efficiencies, and in the resolution
of the \pt of jets and missing transverse momentum ~are corrected by applying scale factors to simulated events, as detailed below.

\section{Event reconstruction and selection}

The particle-flow (PF) algorithm~\cite{CMS-PAS-PFT-09-001,CMS-PAS-PFT-10-001},
which exploits information from all subdetectors, is used to identify
individual particles, such as charged and neutral hadrons, muons,
electrons, and photons. These reconstructed particles are used as input for
reconstructing more complex objects such as \tauh candidates, jets, and variables like missing transverse momentum.

The reconstructed interaction vertex with the largest value of
$\sum_i (\pt^i)^2$,
where $\pt^i$ is the transverse momentum of the $i$th track associated with
the vertex, is selected as the primary vertex of the event. This vertex is used as
the reference vertex for all the objects reconstructed using the PF algorithm.

Electrons are reconstructed by matching the energy deposits in the ECAL to
tracks reconstructed in the silicon pixel and strip detectors.
The electrons selected in this analysis are required to have
transverse momenta $\pt > 50\GeV$~and pseudorapidity~$\abs{\eta}<2.1$~\cite{Khachatryan:2015hwa}.
The identification and isolation of electrons
are based on a multivariate technique~\cite{Hocker:2007ht} and selected electrons
must satisfy tight electron identification and isolation criteria.

Muon reconstruction starts by matching  tracks in the silicon tracker
with tracks in the outer muon spectrometer~\cite{Chatrchyan:2012xi}.
A global muon track is fitted to the hits from both tracks.
Muons are required to have $\pt > 50\GeV$ and
$\abs{\eta}<2.1$.
Quality selection criteria are applied to the muon tracks to distinguish muons originating from particle collisions
with those muons coming from cosmic rays.
In addition, muons are required to pass
isolation criteria to separate  prompt muons from those associated with a
jet, usually from the semileptonic decays of heavy quarks.

The hadron-plus-strips  algorithm~\cite{Khachatryan:2015dfa,CMS-PAS-TAU-16-002}
is used to reconstruct $\tauh$ candidates. It starts from a jet
and searches for candidates produced by the main hadronic decay modes
of a $\tau$ lepton:
either directly to one charged hadron, or via intermediate
$\rho(770)$ and $\mathrm{a_1(1280)}$ mesons to one
charged hadron plus one or two neutral pions, or three charged hadrons.
The reconstructed $\tauh$ is required to have $\abs{\eta}<2.3$ and
$\pt> 50$ ($\pt > 60)$\GeV in the LQ (heavy RH neutrino) search.
Hadronic tau lepton decays are identified by a multivariate technique
that uses as inputs the isolation of the $\tauh$ and
variables that are sensitive to its lifetime.
A selection criterion is used that has an efficiency of approximately 65\% for identifying hadronically decaying tau leptons and a probability of less than 1\% for misidentifying jets as hadronic tau decays.
Additional criteria are applied to remove electrons and
muons reconstructed as $\tauh$ candidates.

The identified electron or muon and the $\tauh$ are required to originate from
the same vertex and be spatially separated by
$\Delta R \equiv \sqrt{\smash[b]{\Delta\phi^2 + \Delta\eta^2} }> 0.5$.
To suppress background events such as diboson and \cPZ+jets with bosons decay giving a final state with a pair of leptons, events containing additional electron or muon candidates with $\pt>15\GeV$,
and which pass loose identification and isolation criteria, are rejected.

Jets are reconstructed using the anti-\kt algorithm
with a distance parameter of $R=0.4$~\cite{Salam:2009jx,Cacciari:2008gp}
using PF candidates. The jet energy is corrected
for the average contribution from particles from other
proton-proton collisions in the same or neighbouring bunch crossings (pileup)~\cite{CMS-PAS-JME-14-001}.
Additional corrections are applied to better reflect the true total momentum of the particles in the jet ~\cite{JME-13-004}.
Selected jets are required to be within $\abs{\eta}<2.4$ and
have $\pt>50$\GeV, and to be separated from the selected
electron or muon and the $\tauh$ by $\Delta R>0.5$.
Further identification requirements are applied to distinguish genuine jets from those coming from pileup~\cite{JME-13-005}.

The transverse momentum imbalance, (\ptvecmiss), is calculated as the negative
vectorial sum of transverse momenta of all PF candidates, and
corrected by propagating the
corrections applied to identified jets~\cite{CMS-PAS-JME-16-004}.
A correction is applied to account for the effect of
additional pileup interactions. In addition, several filters are employed to veto events with large \ptvecmiss caused by detector effects.

Candidate events were collected using a set of triggers
requiring the presence of either an electron or a muon candidate with $\pt>45$\GeV.

The search for LQs is based on a sample of events containing one light lepton,
one $\tauh$ candidate, and
at least two jets.
At least one of the two leading jets is identified as
originating from b quark hadronization (b-tagged) using the
combined secondary vertex  algorithm~\cite{CMS-PAS-BTV-15-001}. The chosen b tagging working point corresponds to an identification efficiency of approximately 70\% with about 1\% misidentification rate from light quarks.
The lepton and $\tauh$ candidate are required to have opposite electric charge.
There are two possible combinations of two tau candidates, with two jets, and the combination that minimises the difference in masses between the two resulting tau candidate-jet systems is chosen.
Additionally,
the invariant mass of the system formed by the visible particles of the $\tauh$ candidate and a jet
is required to be greater than 250\GeV.

The search for a $\PWR$~ boson decaying into a heavy neutrino uses the same data sample as used by the LQ search.
The \ptvecmiss is required to be above 50\GeV and
the invariant mass of the light lepton and the $\tauh$
is required to be greater than 150\GeV.

In the LQ analysis, the fraction of signal events passing all selection cuts ranges
between 1 and 5\% for masses between 300 and 1500\GeV, and
in the $\PWR$ analysis the fraction varies
between 2 and 7\% for masses between 1 and 4\TeV.

The presence of a signal is investigated by analysing
the distribution of \ST. This is defined as the scalar sum of the \pt of the electron or muon, the $\tauh$ candidate, the two jets, and the missing transverse energy.

\section{Background estimation}

Several SM processes can mimic the signatures explored in this search.
Production of  \ttbar pairs is the dominant background because of the presence of  genuine
leptons, \ptvecmiss, and both light- and heavy-flavour jets. Additionally,
the production of a W or Z boson in association with jets,
production of a diboson or a single top quark, and
Quantum ChromoDynamics (QCD) multijet processes can also contribute to the SM background contributions.

Simulated \ttbar events are reweighted according to the top
quark \pt distribution measured in data~\cite{Khachatryan:2015oqa,Chatrchyan:2012saa}.
The normalization and shape of the \ttbar background is then
verified by comparing to a data sample that
consists of events containing an electron, a muon, and at least two jets and including all final selection requirements.
The purity of \ttbar events in this sample exceeds 95\%. Signal contamination in this control region is found to be negligible and does not affect the comparison of data with simulation even in the tail of the $\ST$~distribution.
The normalization and shape of the \ttbar
simulated sample agree well with those observed in data.
Thus, the simulation is used to predict the \ttbar background
in the signal region.

The $\PW$+jets background arises mainly from events
with a genuine electron or muon originating
from the leptonic decay of a $\PW$ boson and an initial- or final-state radiation jet misidentified as a $\tauh$ candidate.
The normalization and shape of the $\PW$ background are obtained from
simulation and a correction factor is applied to the normalization
to take into account differences between data and
simulation.  The W background correction factor is estimated in a data sample that
consists of $\PW\to\mu\nu$ events with three or more jets.
One of the jets is required to pass the \tauh identification criteria.
To reduce the contamination from \ttbar background, events containing jets that pass the b tagging criteria are rejected.
The expected signal contamination in this sample is negligible.
A binned maximum likelihood fit to the transverse mass distribution of the muon and  \ptvecmiss is then performed to derive the W background normalization correction factor.
The transverse mass distribution
is found to have the most discriminating power for separating the $\PW$~background from
the other backgrounds.
As an input to the fit, the normalization and shape of all
other contributions are estimated from simulation. The uncertainties
in the cross sections of all backgrounds are included as nuisance
parameters in the fit.
The contamination from QCD multijet events
is small and derived from simulation.
The best fit value for the W normalization correction factor is found to be $1.0 \pm 0.2$, with the uncertainty including both statistical and systematic components.

 A similar procedure is repeated for the \etau~channel in a control region
containing events with an electron and three or more jets.
The W+jets
normalization factor measured in $\PW\to\Pe\nu$ events is
found to be consistent with the normalization factor
derived in $\PW\to\mu\nu$ events, albeit with a slightly
larger uncertainty.

The contribution of the QCD multijet background to the signal region in both the \mutau
and \etau channels is
estimated from data.
Events in the multijet control region are selected
by inverting the \tauh identification criteria:
the \tauh candidate is required to pass looser, but to fail
tighter, identification criteria.
The events are weighted by the \pt-dependent probability
for a jet satisfying loose isolation criteria to pass the tight
\tauh isolation criteria.
This probability is measured as a function of jet \pt  for \etau and \mutau channels separately, in independent data samples  that are composed of events in which the lepton fails the isolation criteria and has the same charge as \tauh candidate. This probability varies from 20\% for a  jet \pt of 50\GeV to  2\% for a  jet \pt of 400\GeV and is similar for both final states.

In addition to the principal backgrounds, which are estimated as discussed above, other minor backgrounds, arising from single top quark, \cPZ~boson, and diboson production, are estimated from simulation.  The relative contributions of all these backgrounds are given in Tables~\ref{table:lq} and ~\ref{table:rhw}, in Section ~\ref{sec:results}. In these tables, the electroweak background represents the sum of the backgrounds from  \PW+jets,  \cPZ~boson, and diboson production. Additionally, for each channel the background estimation is compared with the observed data and with an estimated representative signal.

\section{Systematic uncertainties}
\label{sec:objectselection}

The results of the analysis are obtained from a binned fit
to the \ST distributions in the \etau and \mutau channels.
Systematic uncertainties may affect the normalization
and/or the shape of the \ST distribution of the signal
and background processes.

The uncertainty in the integrated luminosity of the analysed dataset
amounts to 6.2\%~\cite{CMS-PAS-LUM-15-001}.
Uncertainties in the muon and electron identification and trigger efficiency are determined using
the ``tag-and-probe''
technique~\cite{tag-and-probe}
and amount to 2\% for identification and 5\% for trigger efficiencies.
The $\tauh$ identification efficiency~\cite{TAU-11-001,Khachatryan:2015dfa} is measured in bins of $\tauh$ candidate \pt in $\PZ\to\tau\tau$ and $\PW\to\tau\nu$ events and fitted by a linear function within the range 20 to 200\GeV.
The uncertainty in the $\tauh$ identification efficiency measurement is 6\% for $\tau$ leptons from the decay of Z bosons.
The extrapolation to higher transverse
momenta is taken into account by adding an uncertainty that increases
linearly with \pt and has a value of  20\% for a \pt of 200\GeV.  This uncertainty has a direct effect on the \ST distribution and hence is considered as a shape uncertainty.
Changes in the acceptance due to the uncertainty in the b tagging efficiency and in the mistag rate are measured to be between 3 and 5\%, depending on the process.
The uncertainty in the normalization of the \ttbar~background due to the PDF and scale uncertainties amounts to 5\%~\cite{Czakon:2011xx,Botje:2011sn}.
A 10\% uncertainty is attributed to the Z boson background estimate, while the uncertainty in both the diboson and single top background estimates amounts to 15\%~\cite{Chatrchyan:2013oev}.
The uncertainty in the yield of QCD multijet and $\PW$+jet
backgrounds amounts to 30\%.
The uncertainty in the signal acceptance due to the choice of the PDF set
in the simulated sample is evaluated in accordance to the PDF4LHC
recommendations~\cite{Alekhin:2011sk, Botje:2011sn},
by comparing the results obtained using the
CTEQ6.6L, MSTW08, and NNPDF10  PDF
sets~\cite{Nadolsky:2008zw,Martin:2007bv,Ubiali:2008uk}
with those from the default PDF set (CTEQ6L1).
This uncertainty amounts to 5\%~\cite{CMS-PAS-EXO-16-016}.

The energy scales (ES) of the $\tauh$ candidate and the associated jet affect the shape of
the \ST distribution and normalization of the signal and
background processes.
The effects of ES uncertainties on the analysis are estimated by
varying the \tauh and jet energies within their respective
uncertainties and recomputing \ST after the final selection.
The uncertainty in the \tauh ES amounts to 3\%~\cite{Khachatryan:2015dfa}.
The uncertainty in the jet ES affects the \pt spectrum of the jets and consequently \ptvecmiss, and
is applied to signal and all backgrounds that are estimated with MC simulation~\cite{Chatrchyan:2011ds}.
The uncertainties in the electron, muon, and \ptvecmiss ES have a negligible effect on the \ST distribution. The uncertainty in the top quark
\pt reweighting correction is derived by changing the event weight between zero and twice the nominal reweighting correction
value~\cite{Khachatryan:2015oqa,Chatrchyan:2012saa}. All these three uncertainties are treated as correlated between the \etau and \mutau channels.

Finally, the effects of statistical uncertainties associated with the signal and background shapes or with the numbers of events in the data control regions are included in the analysis. The statistical uncertainties are uncorrelated across the bins in each background distribution~\cite{BarlowBeeston}.

Systematic and statistical uncertainties are
represented by nuisance parameters in the fit.
A log-normal probability distribution function is assumed
for the nuisance parameters that affect the event yields
of the various background contributions. Systematic
uncertainties affecting the  $\ST$ distributions are assumed
to have a Gaussian probability distribution function. Among those uncertainties, the \tauh ES and high \pt \tauh extrapolation uncertainties are uncorrelated between the \etau and  \mutau channels, because of the different \tauh identification criteria used to reduce the electron and muon mis-identification rate in each channel.
The jet ES is treated as correlated across the two channels.

\section{Results}
\label{sec:results}

\begin{figure}
  \centering
    \includegraphics[width=\cmsFigWidth]{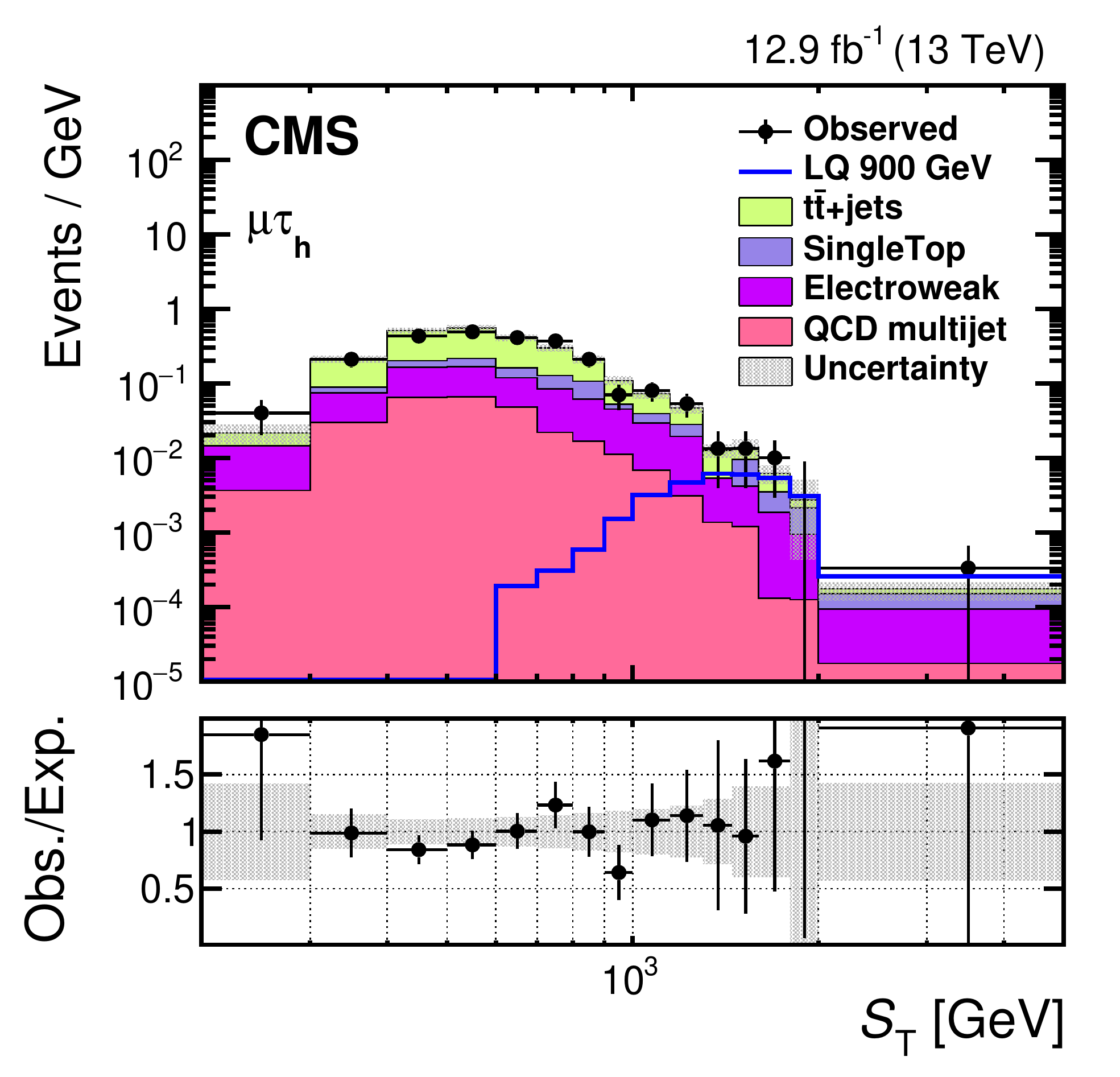}
    \includegraphics[width=\cmsFigWidth]{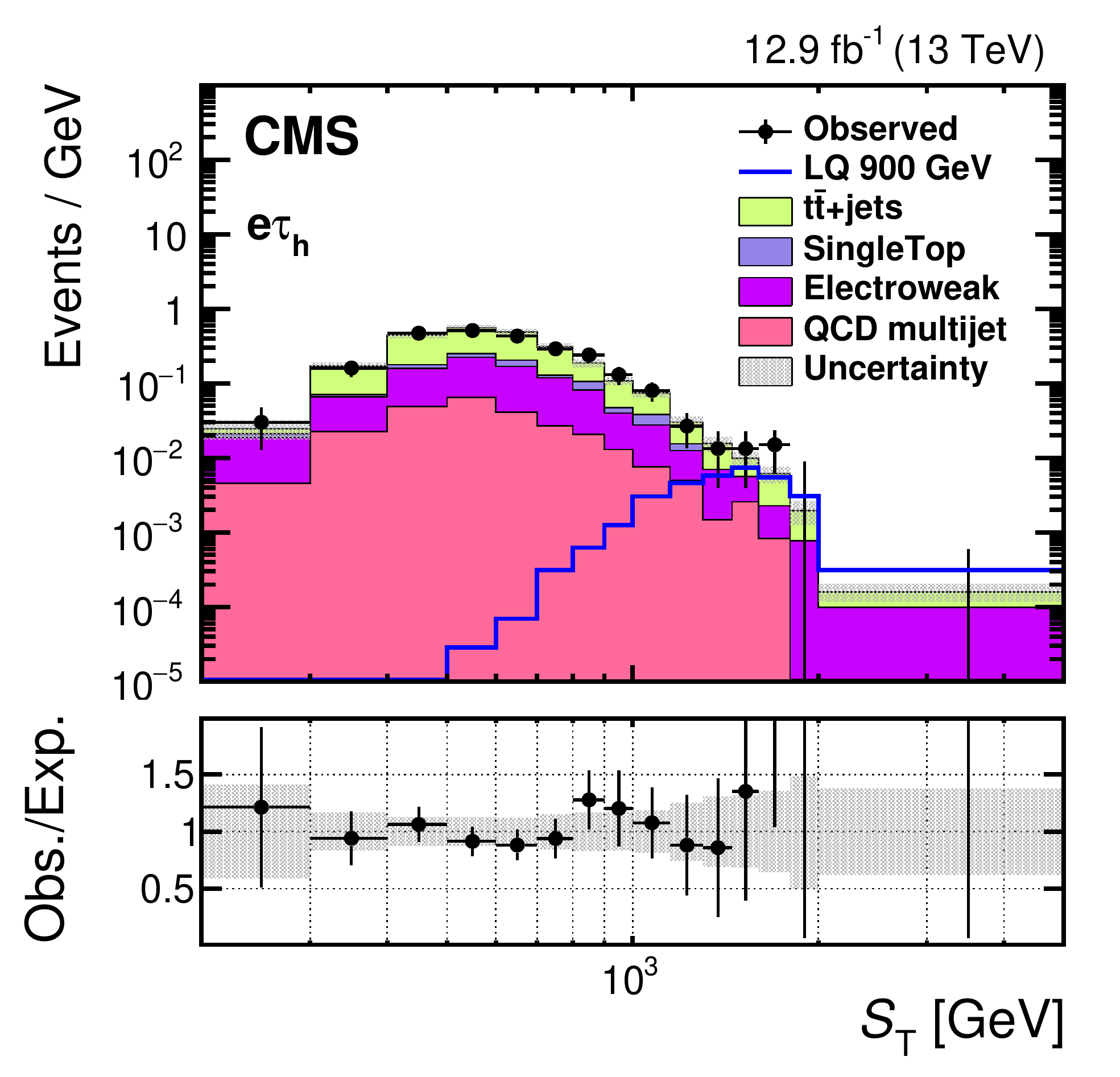}
    \includegraphics[width=\cmsFigWidth]{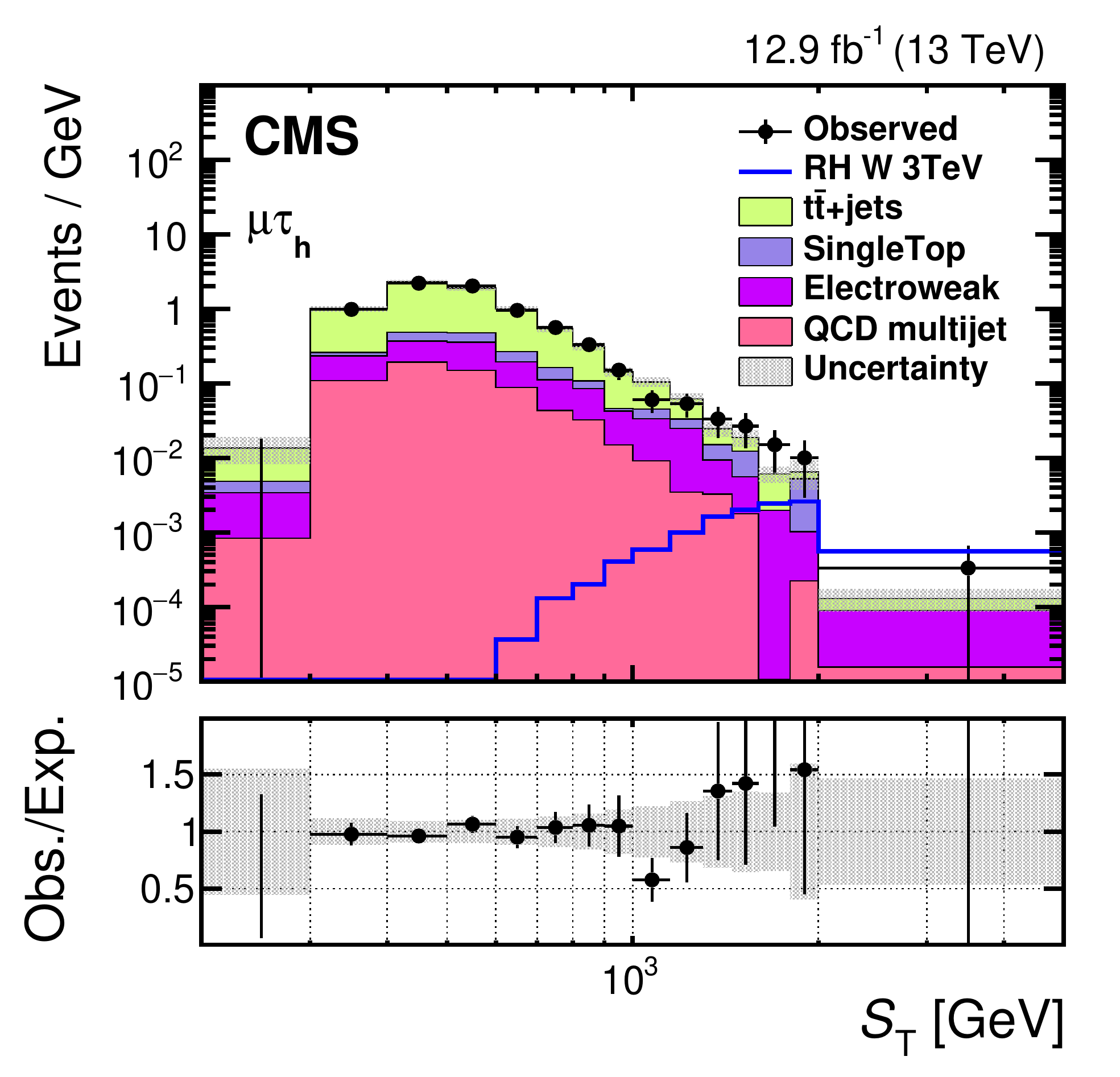}
    \includegraphics[width=\cmsFigWidth]{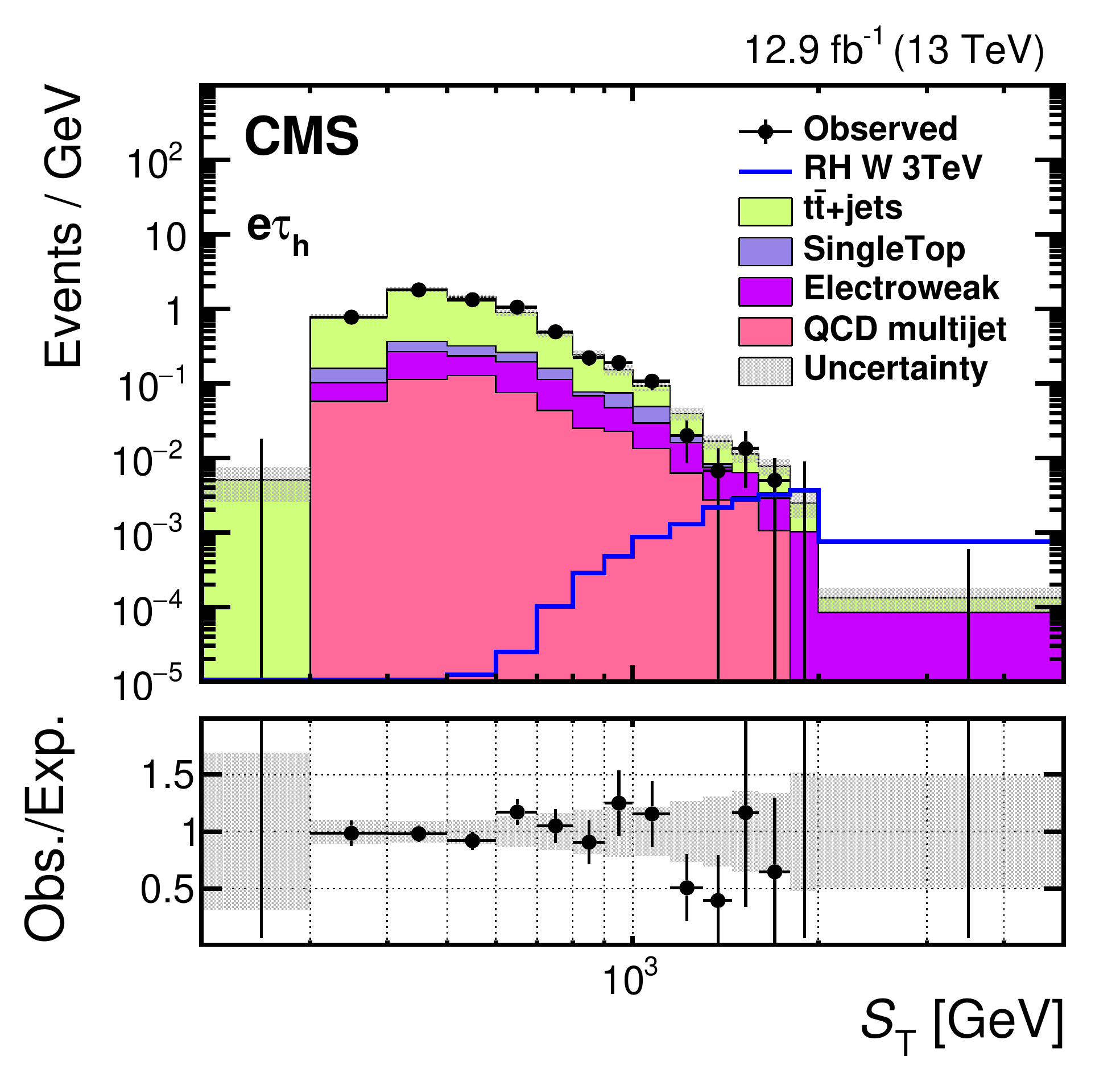}
    \caption{Measured \ST distribution in the \mutau (left) and \etau (right) channels of the LQ (upper) and heavy RH neutrino (lower) analyses, compared to the expected SM background contribution. A hypothetical LQ signal of mass ${M_\mathrm{LQ}=900}$\GeV and a hypothetical heavy $\PWR$ signal of mass ${M}_{\PWR}=3$\TeV are overlaid to illustrate the sensitivity. The electroweak background represents the sum of \PW~boson, \cPZ~boson, and diboson production. The last bin of each plot contains overflow events. A binned maximum likelihood fit is performed on the \ST distribution. The uncertainty bands represent the sum in quadrature of statistical and systematic uncertainties, obtained from the fit. The lower panels in all plots compare the observed and expected events in each bin.}
    \label{fig:STplotsLQ}
\end{figure}

A binned maximum likelihood fit to the $\ST$
distribution has been applied to the  \etau and \mutau channels
simultaneously. The signal production rate is constrained to the same value in the two channels.
The \ST distributions for both the LQ and $\PWR$ analyses are shown in Fig.~\ref{fig:STplotsLQ}. Shape,  normalization and uncertainty are shown for the values of nuisances parameters obtained from the fit.
No excess is seen above the SM expectation within the statistical and systematic uncertainties in both searches.
The event yields observed in the leptoquark analysis and in the heavy right-handed W boson analysis are shown in Tables~\ref{table:lq} and ~\ref{table:rhw}, respectively, and compared to background expectations and to the estimated event yields for representative signals.

\begin{figure}
\centering
 \includegraphics[width=\cmsFigWidth]{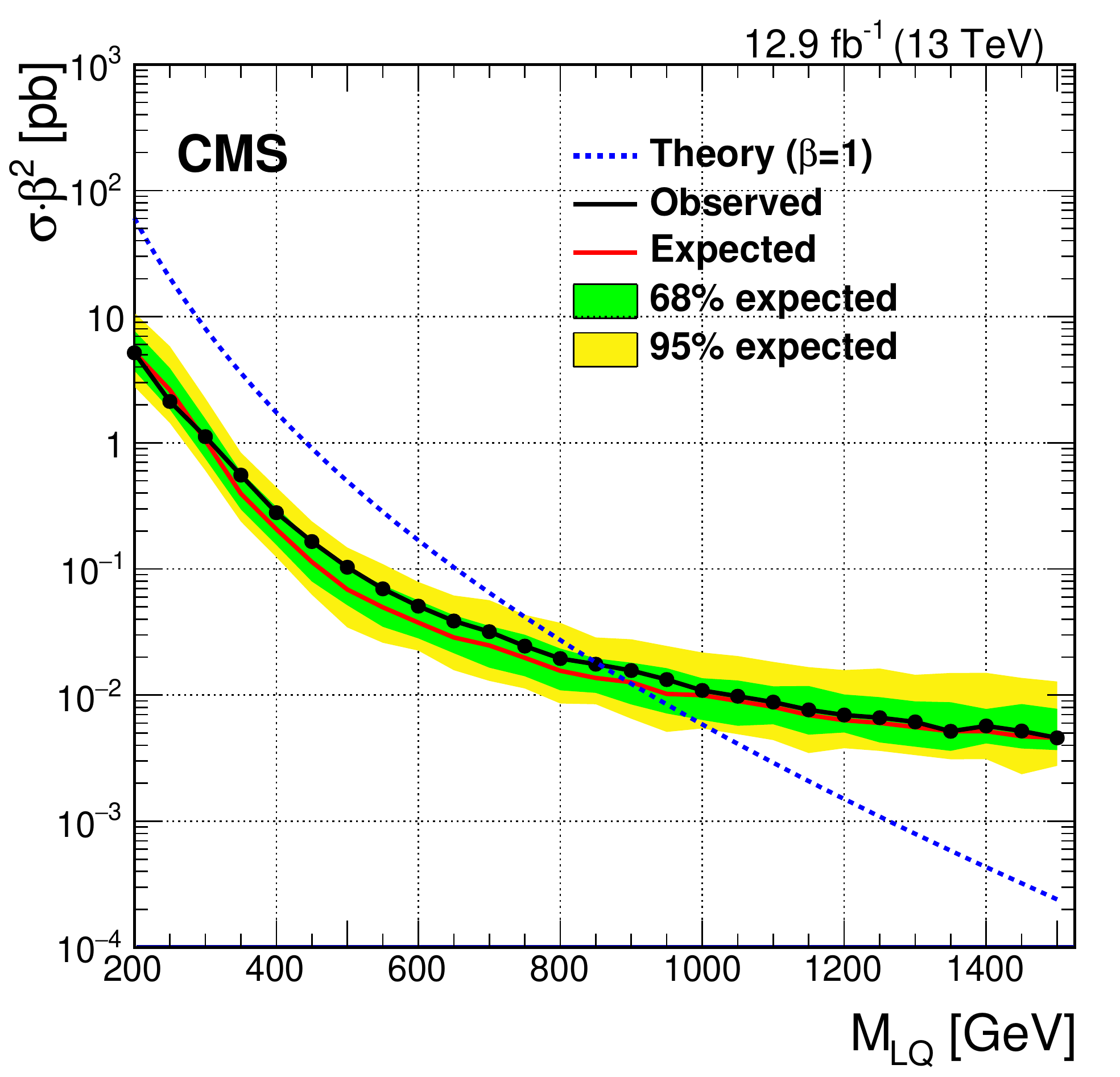}
  \includegraphics[width=\cmsFigWidth]{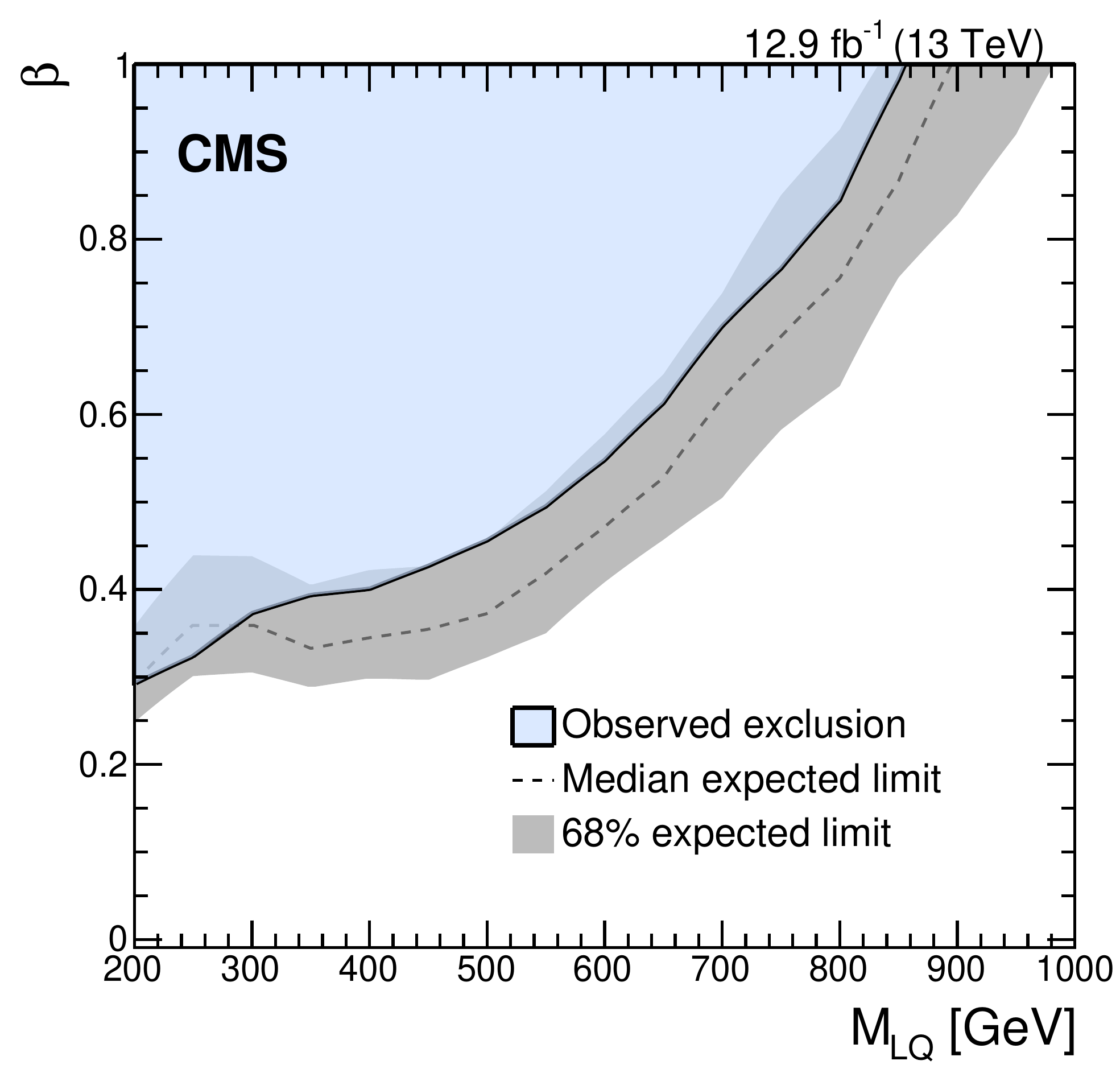}
    \caption{Observed and expected limits at 95\% CL on the product of cross section and branching fraction squared, obtained from the combination of the \etau and \mutau channels, in the LQ  analysis (left) and 95\% CL observed and expected exclusion limits on the LQ mass, as a function of $\beta$(right).
    In the left plot, the green and yellow bands represent the one and two standard deviation uncertainties in the expected limits. The dashed dark blue curve represents the theoretical LQ pair production cross section, assuming $\beta = 100$\%~\cite{Kramer:2004df,Kramer:1997hh}. In the right plot, the grey band represents the one standard deviation uncertainty in the expected limit.}
    \label{fig:LimitplotsLQ}
\end{figure}

Upper limits on the product of the cross section and
branching fractions are set at  95\% CL using a modified frequentist criterion CL$_\mathrm{s}$~\cite{CLS2,Read:2002hq}, based on
the binned distribution of the
$\ST$ variable.
Figure~\ref{fig:LimitplotsLQ} (left) shows the observed and expected  95\% CL upper limit on the product of cross section and branching fraction in the LQ analysis.
The observed (expected) 95\% CL mass limit for third-generation scalar LQ is determined to be 850 (900)\GeV, respectively,
assuming $\beta = 1$, namely a 100\% branching fraction for the LQ to decay into a $\tau$ lepton and a bottom quark.
Figure~\ref{fig:LimitplotsLQ} (right)
shows the 95\% CL observed and expected exclusion limits on the LQ mass, as a function of $\beta$.

\begin{figure}
  \centering
\includegraphics[width=\cmsFigWidth]{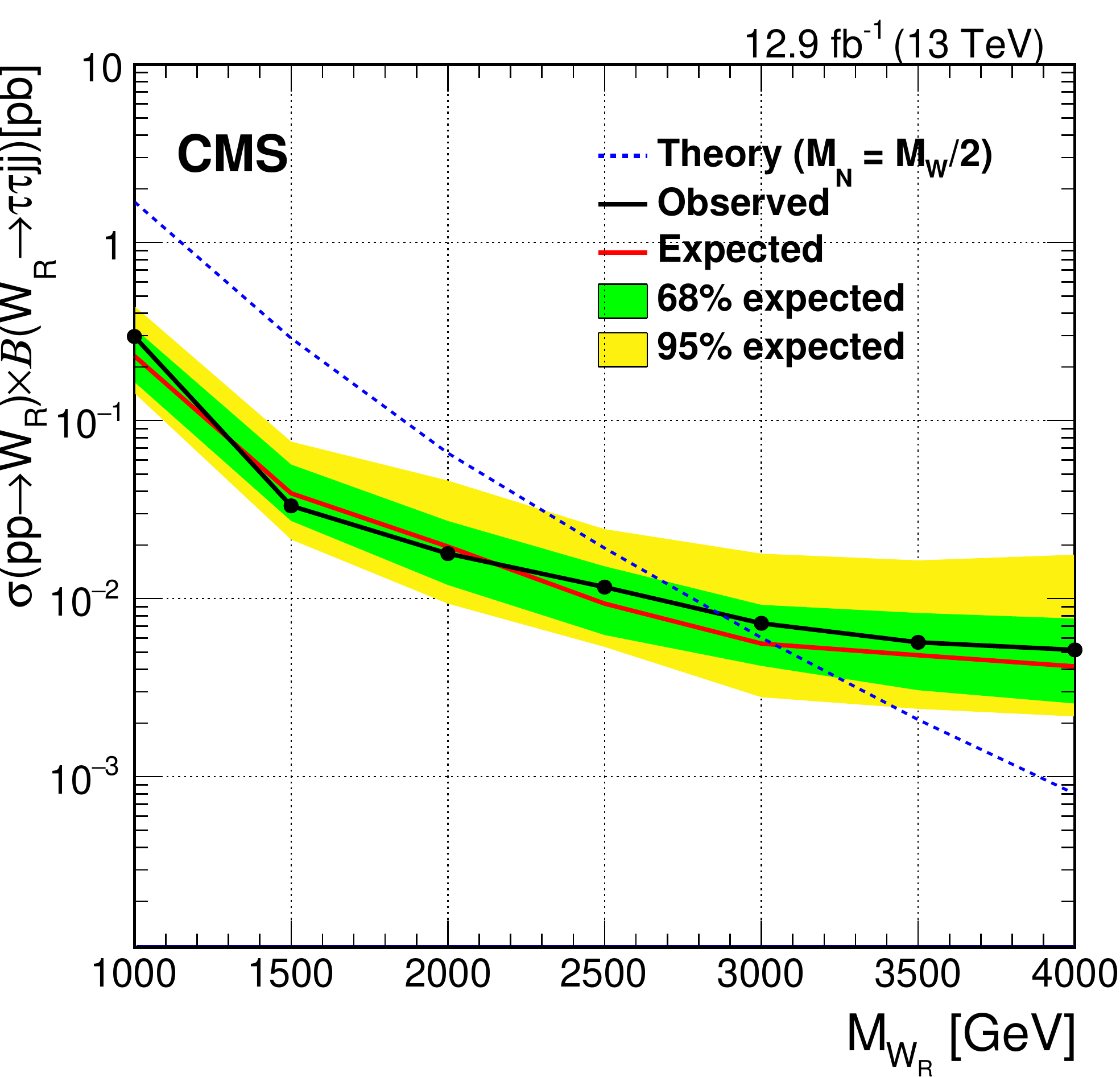}
\includegraphics[width=\cmsFigWidth]{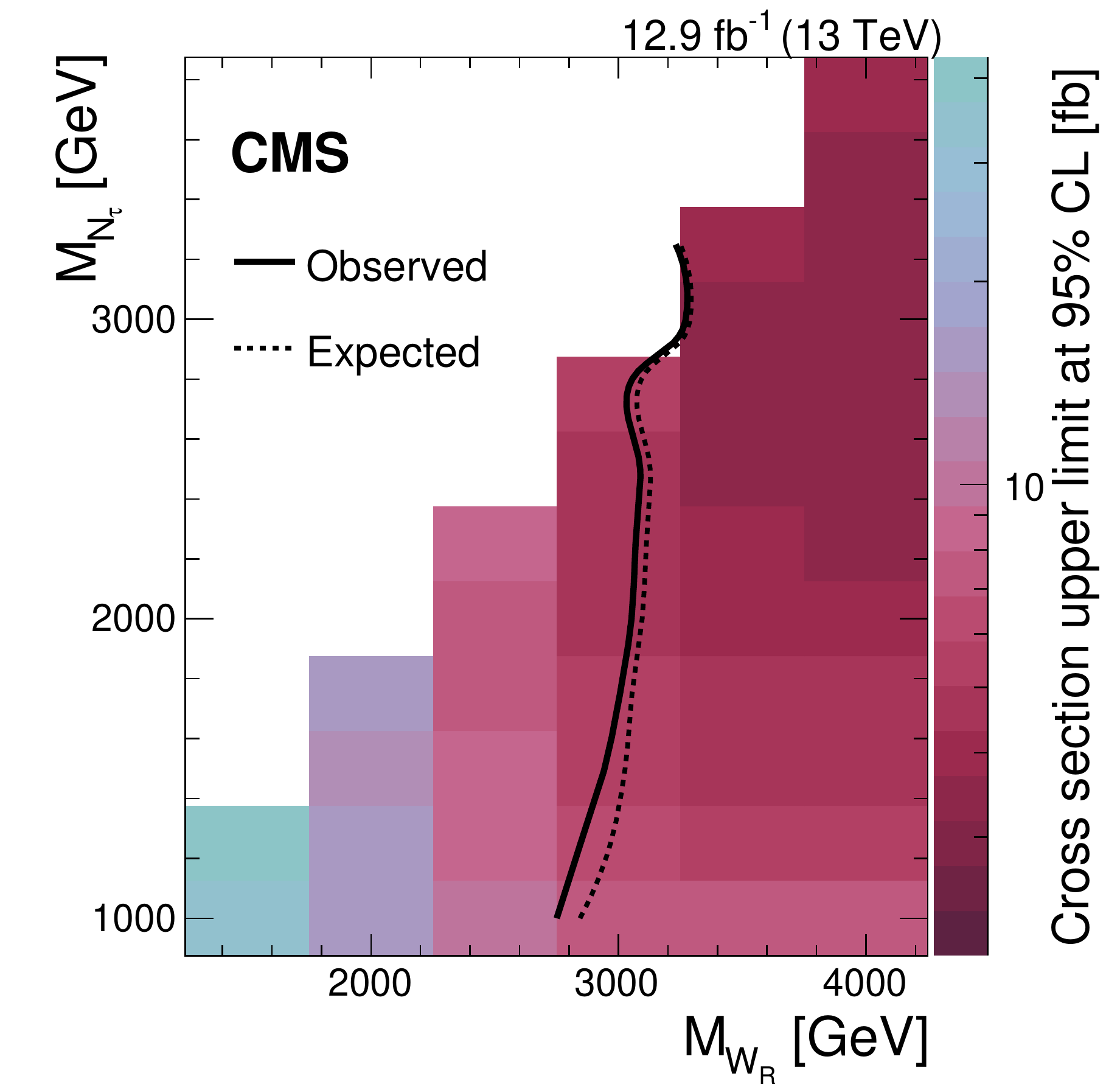}
\caption{Observed and expected  limits at  95\% CL on the product of cross section and branching fraction, obtained from the combination of the \etau and \mutau channels in the heavy right-handed neutrino analysis (left) and the observed and expected  limits at  95\% CL on the production cross section as a function of $\mathrm{M}_{\PWR}$ and $\mathrm{M}_{\NR}$ (right).
     The green and yellow bands represent the one and two standard deviation uncertainties in the expected limits.
     The dashed dark blue curve represents the theoretical prediction for the product of the $\PWR$ boson production cross section and the branching fraction for decay to a $\tau$  lepton and RH neutrino, assuming the mass of the RH neutrino to be half the mass of the $\PWR$ boson~\cite{delAguila:2007qnc}.}
    \label{fig:LimitplotsRH}
\end{figure}

Figure~\ref{fig:LimitplotsRH} (left) shows the observed and expected
upper limits at 95\% CL on the product of cross section and
branching fraction for the $\PWR\to\tau\Nt$ analysis.
Assuming the mass of the neutrino
to be half the mass of the $\PWR$ boson,
the observed (expected) limit at 95\% CL
on the mass of heavy right-handed $\PWR$ bosons
is determined to be 2.9 (3.0)\TeV, respectively.
Figure~\ref{fig:LimitplotsRH} (right) shows the observed and
expected 95\% CL upper limits on the production cross section
as functions of ${M}_{\PWR}$ and ${M}_{\Nt}$.
The blue curve in the left plot represent the theoretical production cross section of $\PWR$ boson times branching fraction of the $\PWR$ boson to a $\tau$ lepton and RH neutrino, assuming mass of  RH neutrino to be half the mass of $\PWR$ boson.

\begin{table}[htbp]
\small
\caption{Number of events observed in the $\etau$ and $\mutau$ channels of the LQ analysis compared to the background expectations and to the event yield expected for a representative LQ signal of mass 900 GeV. The quoted uncertainties represent the sum of statistical and systematic uncertainties and are obtained by the binned maximum likelihood fit of the distribution in \ST, as described in the text.}

\begin{center}
\begin{tabular}{|l|r@{$ \,\,\pm\,\, $}lr@{$ \,\,\pm\,\, $}l|}
\hline
Process & \multicolumn{2}{c}{e$\tauh$} & \multicolumn{2}{c|}{$\mu\tauh$} \\
\hline
\hline
\ttbar+jets                & $136.8$ & $13.2$ & $145.6$ & $14.6$ \\
SingleTop                             & $15.7$ & $3.1$ & $27.6$ & $4.5$ \\
Electroweak                                        & $69.6$ & $10.5$ & $53.3$ & $9.0$ \\
QCD multijet                                        & $25.0$ & $6.8$ & $28.9$ & $7.4$ \\
\hline
Total expected background                        & $247.1$ & $16.8$ & $255.4$ & $16.1$ \\
\hline
LQ 900 GeV                                         & $6.0$ & $0.3$ & $5.7$ & $0.2$ \\
\hline
Observed data                                    & \multicolumn{2}{c}{$249$} & \multicolumn{2}{c|}{$250$} \\
\hline
\end{tabular}

\label{table:lq}
\end{center}
\end{table}

\begin{table}[htbp]
\caption{Number of events observed in the $\etau$ and $\mutau$ channels of the heavy right-handed W analysis compared to the background expectations and to the event yield expected for a representative right-handed W boson signal of mass 3 TeV. The quoted uncertainties represent the sum of statistical and systematic uncertainties and are obtained by the binned maximum likelihood fit of the distribution in \ST, as described in the text. }
\begin{center}
\begin{tabular}{|l|r@{$ \,\,\pm\,\, $}lr@{$ \,\,\pm\,\, $}l|}
\hline
Process & \multicolumn{2}{c}{e$\tauh$} & \multicolumn{2}{c|}{$\mu\tauh$} \\
\hline
\hline
\ttbar+jets                & $456.2$ & $25.8$ & $557.6$ & $30.7$ \\
SingleTop                             & $41.6$ & $7.1$ & $47.6$ & $8.1$ \\
Electroweak                                        & $60.2$ & $26.8$ & $83.6$ & $33.2$ \\
QCD multijet                                        & $48.6$ & $13.1$ & $65.7$ & $16.8$ \\
\hline
Total expected background                        & $606.6$ & $33.5$ & $754.5$ & $38.7$ \\
\hline
RHW 3000 GeV                                         & $4.8$ & $0.3$ & $3.5$ & $0.3$ \\
\hline
Observed data                                    & \multicolumn{2}{c}{$606$} & \multicolumn{2}{c|}{$751$} \\
\hline
\end{tabular}
\label{table:rhw}
\end{center}
\end{table}

\section{Summary}
Searches have been performed for third-generation scalar leptoquarks
and for heavy right-handed neutrinos
in events containing one electron or muon, one hadronically decaying $\tau$ lepton,
and two or more jets,
using  pp collision data at $\sqrt{s} = 13$\TeV, recorded
by the CMS detector at the LHC and
corresponding to an integrated luminosity of 12.9\fbinv.
The data are found to be in good
agreement with the standard model prediction in both analyses. A limit at 95\% confidence level is set
on the product of the leptoquark pair production cross section and $\beta^2$, where
$\beta$ denotes the branching fraction for the decay of the leptoquark into a $\tau$
lepton and a bottom quark. Assuming $\beta=1$, third-generation leptoquarks with masses
below 850\GeV are excluded at 95\% confidence level.
In the heavy RH neutrino analysis,
considering the decay $\PWR\to\tau\NR$ and assuming the mass of the
heavy neutrino to be half the mass of the $\PWR$ boson, we exclude
$\PWR$ boson masses below 2.9\TeV at 95\% confidence level.
These are the best mass limits to date for third-generation leptoquarks and heavy right-handed neutrinos with $\tau$ leptons in the final state.

\begin{acknowledgments}
We congratulate our colleagues in the CERN accelerator departments for the excellent performance of the LHC and thank the technical and administrative staffs at CERN and at other CMS institutes for their contributions to the success of the CMS effort. In addition, we gratefully acknowledge the computing centres and personnel of the Worldwide LHC Computing Grid for delivering so effectively the computing infrastructure essential to our analyses. Finally, we acknowledge the enduring support for the construction and operation of the LHC and the CMS detector provided by the following funding agencies: BMWFW and FWF (Austria); FNRS and FWO (Belgium); CNPq, CAPES, FAPERJ, and FAPESP (Brazil); MES (Bulgaria); CERN; CAS, MoST, and NSFC (China); COLCIENCIAS (Colombia); MSES and CSF (Croatia); RPF (Cyprus); SENESCYT (Ecuador); MoER, ERC IUT, and ERDF (Estonia); Academy of Finland, MEC, and HIP (Finland); CEA and CNRS/IN2P3 (France); BMBF, DFG, and HGF (Germany); GSRT (Greece); OTKA and NIH (Hungary); DAE and DST (India); IPM (Iran); SFI (Ireland); INFN (Italy); MSIP and NRF (Republic of Korea); LAS (Lithuania); MOE and UM (Malaysia); BUAP, CINVESTAV, CONACYT, LNS, SEP, and UASLP-FAI (Mexico); MBIE (New Zealand); PAEC (Pakistan); MSHE and NSC (Poland); FCT (Portugal); JINR (Dubna); MON, RosAtom, RAS, RFBR and RAEP (Russia); MESTD (Serbia); SEIDI, CPAN, PCTI and FEDER (Spain); Swiss Funding Agencies (Switzerland); MST (Taipei); ThEPCenter, IPST, STAR, and NSTDA (Thailand); TUBITAK and TAEK (Turkey); NASU and SFFR (Ukraine); STFC (United Kingdom); DOE and NSF (USA).

\hyphenation{Rachada-pisek} Individuals have received support from the Marie-Curie programme and the European Research Council and EPLANET (European Union); the Leventis Foundation; the A. P. Sloan Foundation; the Alexander von Humboldt Foundation; the Belgian Federal Science Policy Office; the Fonds pour la Formation \`a la Recherche dans l'Industrie et dans l'Agriculture (FRIA-Belgium); the Agentschap voor Innovatie door Wetenschap en Technologie (IWT-Belgium); the Ministry of Education, Youth and Sports (MEYS) of the Czech Republic; the Council of Science and Industrial Research, India; the HOMING PLUS programme of the Foundation for Polish Science, cofinanced from European Union, Regional Development Fund, the Mobility Plus programme of the Ministry of Science and Higher Education, the National Science Center (Poland), contracts Harmonia 2014/14/M/ST2/00428, Opus 2014/13/B/ST2/02543, 2014/15/B/ST2/03998, and 2015/19/B/ST2/02861, Sonata-bis 2012/07/E/ST2/01406; the National Priorities Research Program by Qatar National Research Fund; the Programa Clar\'in-COFUND del Principado de Asturias; the Thalis and Aristeia programmes cofinanced by EU-ESF and the Greek NSRF; the Rachadapisek Sompot Fund for Postdoctoral Fellowship, Chulalongkorn University and the Chulalongkorn Academic into Its 2nd Century Project Advancement Project (Thailand); and the Welch Foundation, contract C-1845. \end{acknowledgments}

\bibliography{auto_generated}
\cleardoublepage \appendix\section{The CMS Collaboration \label{app:collab}}\begin{sloppypar}\hyphenpenalty=5000\widowpenalty=500\clubpenalty=5000\input{EXO-16-023-authorlist.tex}\end{sloppypar}
\end{document}

%% file: EXO-16-023-authorlist.tex
\textbf{Yerevan Physics Institute,  Yerevan,  Armenia}\\*[0pt]
A.M.~Sirunyan, A.~Tumasyan
\vskip\cmsinstskip
\textbf{Institut f\"{u}r Hochenergiephysik,  Wien,  Austria}\\*[0pt]
W.~Adam, E.~Asilar, T.~Bergauer, J.~Brandstetter, E.~Brondolin, M.~Dragicevic, J.~Er\"{o}, M.~Flechl, M.~Friedl, R.~Fr\"{u}hwirth\cmsAuthorMark{1}, V.M.~Ghete, C.~Hartl, N.~H\"{o}rmann, J.~Hrubec, M.~Jeitler\cmsAuthorMark{1}, A.~K\"{o}nig, I.~Kr\"{a}tschmer, D.~Liko, T.~Matsushita, I.~Mikulec, D.~Rabady, N.~Rad, B.~Rahbaran, H.~Rohringer, J.~Schieck\cmsAuthorMark{1}, J.~Strauss, W.~Waltenberger, C.-E.~Wulz\cmsAuthorMark{1}
% ...
\vskip\cmsinstskip
% ...
1:~~Also at Vienna University of Technology, Vienna, Austria\\
% ...
